\documentclass[twocolumn, superscriptaddress,amsmath, prl, aps]{revtex4-2}

\usepackage[T1]{fontenc}
\usepackage[utf8]{inputenc}
\usepackage{siunitx}
\usepackage{times,color,amsthm,graphics,graphicx,bm,bbm,dcolumn}
\usepackage{epsfig}
\usepackage{graphicx}
\usepackage{xcolor}
\usepackage[colorlinks,urlcolor=black,citecolor=blue,linkcolor=black]{hyperref}
\usepackage{soul,bm}
\usepackage{verbatim}
\usepackage{orcidlink}

\begin{document}
\setlength{\tabcolsep}{18pt}


\setcounter{equation}{0}
\setcounter{figure}{0}
\setcounter{table}{0}

\clearpage
\setcounter{page}{1}

\title[Article Title]{Measuring mutual friction in superfluids: the role of initial vortex configuration fluctuations}
\date{\today}

\author{N.~Grani~\orcidlink{0000-0001-6107-9726}}
\email[E-mail: ]{grani@lens.unifi.it}
\affiliation{Department of Physics, University of Florence, 50019 Sesto Fiorentino, Italy}
\affiliation{European Laboratory for Nonlinear Spectroscopy (LENS), University of Florence, 50019 Sesto Fiorentino, Italy}
\affiliation{Istituto Nazionale di Ottica del Consiglio Nazionale delle Ricerche (CNR-INO) c/o LENS, 50019 Sesto Fiorentino, Italy}
\affiliation{INFN, Sezione di Firenze, 50019 Sesto Fiorentino, Italy}

\author{D.~Hern\'andez-Rajkov~\orcidlink{0009-0002-1908-4227}}
\affiliation{European Laboratory for Nonlinear Spectroscopy (LENS), University of Florence, 50019 Sesto Fiorentino, Italy}
\affiliation{Istituto Nazionale di Ottica del Consiglio Nazionale delle Ricerche (CNR-INO) c/o LENS, 50019 Sesto Fiorentino, Italy}
\affiliation{INFN, Sezione di Firenze, 50019 Sesto Fiorentino, Italy}
\altaffiliation{These authors contributed equally to this work.}

\author{M.~Fr\'ometa Fern\'andez~\orcidlink{0000-0003-4937-8306}}
\affiliation{European Laboratory for Nonlinear Spectroscopy (LENS), University of Florence, 50019 Sesto Fiorentino, Italy}
\affiliation{Istituto Nazionale di Ottica del Consiglio Nazionale delle Ricerche (CNR-INO) c/o LENS, 50019 Sesto Fiorentino, Italy}
\affiliation{INFN, Sezione di Firenze, 50019 Sesto Fiorentino, Italy}

\author{G.~Del Pace \orcidlink{0000-0002-0882-2143}}
\affiliation{Department of Physics, University of Florence, 50019 Sesto Fiorentino, Italy}
\affiliation{European Laboratory for Nonlinear Spectroscopy (LENS), University of Florence, 50019 Sesto Fiorentino, Italy}
\affiliation{Istituto Nazionale di Ottica del Consiglio Nazionale delle Ricerche (CNR-INO) c/o LENS, 50019 Sesto Fiorentino, Italy}
\affiliation{INFN, Sezione di Firenze, 50019 Sesto Fiorentino, Italy}

\author{G.~Roati \orcidlink{0000-0001-8749-5621}}
\affiliation{European Laboratory for Nonlinear Spectroscopy (LENS), University of Florence, 50019 Sesto Fiorentino, Italy}
\affiliation{Istituto Nazionale di Ottica del Consiglio Nazionale delle Ricerche (CNR-INO) c/o LENS, 50019 Sesto Fiorentino, Italy}
\affiliation{INFN, Sezione di Firenze, 50019 Sesto Fiorentino, Italy}

\begin{abstract}
\noindent 
The physical origin of mutual friction in quantum fluids is deeply connected to the fundamental nature of superfluidity. It stems from the interaction between the superfluid and normal components, mediated by the dynamics of quantized vortices that induce the exchange of momentum and energy. Despite the complexity of these interactions, their essential features can be effectively described by the dissipative point vortex model — an extension of classical vortex dynamics that incorporates finite-temperature dissipation.
Mutual friction is parametrized by the longitudinal (dissipative) coefficient $\alpha$ and the transverse (reactive) coefficient $\alpha'$. Accurate measurement of these parameters provides critical insights into the microscopic mechanisms governing vortex motion and dissipation in quantum fluids, serving as a key benchmark for theoretical models. In this work, we employ the dissipative point vortex model to study how fluctuations in the initial conditions influence the inference of $\alpha$ and $\alpha'$ from the time evolution of the vortex trajectories. Using experimentally realistic parameters, we show that fluctuations can introduce significant biases in the extracted values of the mutual friction coefficients. We compare our findings with recent experimental measurements in strongly interacting atomic superfluids.
Applying this analysis to our recent experimental results allowed us to account for fluctuations in the correct determination of $\alpha$ and $\alpha'$.
\end{abstract}

\maketitle

\section{Introduction}

The motion of quantized vortices in superfluids and superconductors is intrinsically linked to dissipation. At finite temperatures, thermal excitations scatter off vortex cores, generating mutual friction between the normal and superfluid components and thereby introducing dissipation into the superflow. Conversely, mutual friction plays a central role in shaping vortex dynamics.
As a result, precise measurements and analyses of vortex motion serve as powerful probes of the underlying microscopic dissipation mechanisms, providing valuable benchmarks for theoretical models of quantum fluid dynamics \cite{Sonin97,kopnin2002vortex}.  

Recent experiments have demonstrated the precise tracking of individual vortex motion in ultracold gases as well as in $^4$He superfluids \cite{moon2015thermal, minowa2025direct, tang2023imaging, kwon2021sound, hernandez2024connecting}. Ultracold atomic gases, in particular, offer a powerful platform for studying vortex dynamics: they enable the precise initialization of arbitrary vortex configurations \cite{Samson_DeterministicCreation,park2018critical,grani2025mutual,kwon2021sound, hernandez2024connecting,kwon2014relaxation,ku2014motion,neely2024melting,simjanovski2024shear,kim2016role}, provide tunable superfluid geometries and controlled interparticle interactions, and allow for direct, high-resolution imaging of the superfluid density, where individual vortices appear as distinct density depletions in time-of-flight images.
However, standard absorption imaging, typically used in ultracold atoms experiments, is inherently destructive, so that vortex trajectories must necessarily be reconstructed from multiple experimental realizations. Each realization is prepared with nominally identical initial conditions, but is probed at different evolution times. While vortex positions can be initialized with a precision comparable to the vortex core size—set by the superfluid healing length—even small fluctuations in the initial configuration can lead to substantial deviations in the long-time evolution of the trajectories.

In this work, we examine how uncertainties in the initial vortex positions influence the observed vortex dynamics. We focus on the specific configuration employed in our recent study \cite{grani2025mutual}, consisting of a vortex–anti-vortex pair in a disk-shaped superfluid, with one vortex pinned at the center. Our analysis considers typical experimental parameters and reveals that initial fluctuations could lead to systematic biases in the extracted mutual friction coefficients, $\alpha$ and $\alpha'$. We identify the origin of these biases and outline strategies to detect their presence, enabling the extraction of effective mutual friction values under realistic experimental conditions.

\begin{figure*}[th]
    \centering
    \includegraphics[]{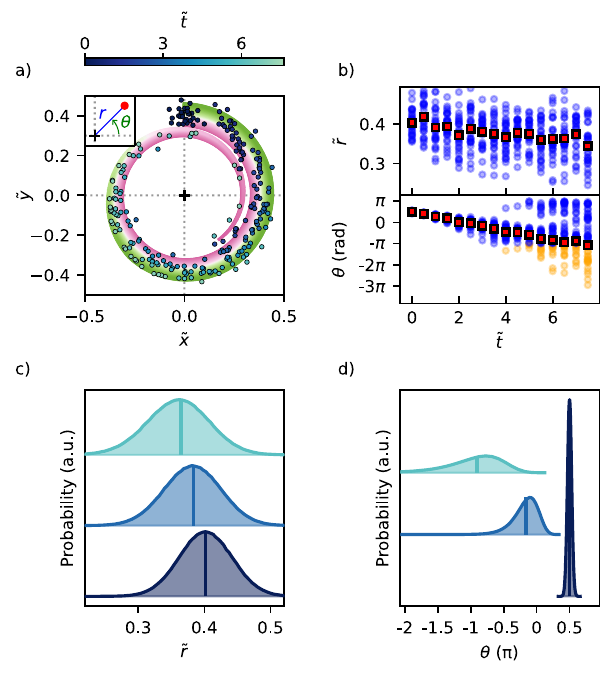}
    \caption{a) DPVM time evolution of vortices characterized by initial conditions randomly distributed with standard deviation in the initial position $\sigma[\tilde{x}_0] = \sigma[\tilde{y}_0]= 0.04$, $\alpha = \mathrm{0.02}$, $\alpha '= \mathrm{0.2}$,
    and $\tilde{r}_0 = 0.4$. Different colors mark different evolution times from dark to light blue. 
    Green to pink shadow represents the evolution of trajectories with different initial radius within the standard deviation up to $\tilde{t} = 7.7$. 
    b) Time evolution of the radial and angular position of the point displayed in a). Blue points (red squares) are the observed value for each realization (average value) following the analysis procedure applied to the experimental data. Yellow points represent the true values as obtained from Eq.~\eqref{eq:theta}, when these differ from the observed one.
    Error bars for the red points are smaller than the symbols and represent the standard deviation of the mean.
    c)-d) Probability distribution of the radial (c) and angular (d) positions obtained from the evolution of $10^6$ different initial conditions as in a) for $\tilde{t} = 0$, dark navy (bottom), $\tilde{t} = 3$, medium blue (middle), and $\tilde{t} = 6$, light cyan (top). Color code is the same as a).
    Distributions at different times are vertically shifted to increase visibility.
    Vertical lines with same colors mark the time evolution of the average value of the initial distributions.}
    \label{fig1}
\end{figure*}

\begin{figure*}[t!]
\centering
\includegraphics[]{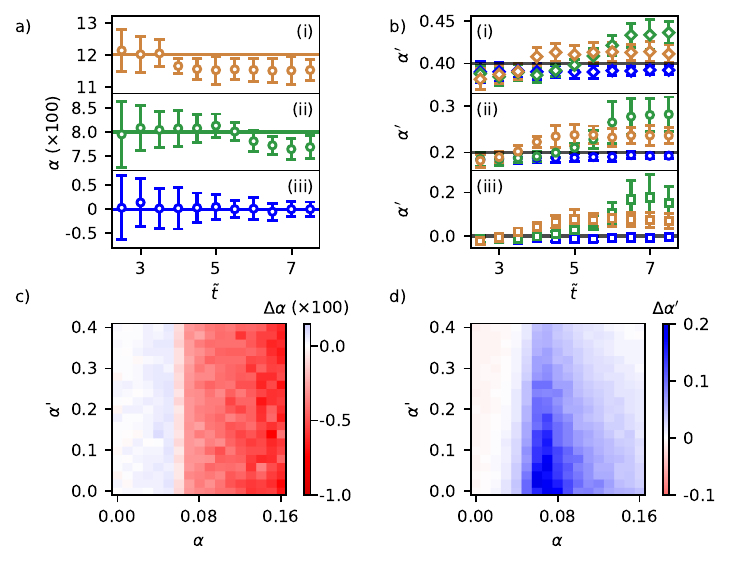}
\caption{
a-b) Extracted values of $\alpha$ (a) for real values of $\alpha = 0.12$ (i), $0.08$ (ii) and $0$ (iii) and  $\alpha'$ (b) for real values of $\alpha ' = 0.4$ (i), $0.2$ (ii) and $0$ (iii). 
Simulations were carried out with $\tilde{r}_0 = 0.4$, $\sigma[\tilde{x}_0] = \sigma[\tilde{y}_0] = 0.03$, $\Delta \tilde{t} = 0.25$ and $\tilde{r}_{min} = 0$. 
Blue, green and gold markers correspond to simulations with true values of $\alpha = 0.00$, $0.08$, and $0.12$, respectively. Squares, circles and diamonds correspond to a real value of $\alpha' = 0.0$, $0.2$ and $0.4$, respectively.
Horizontal solid lines indicate the true input values.
c)-d) Deviation in the determination of the mutual friction coefficients $\Delta \alpha$ and $\Delta \alpha '$, as a function of the real values at $\tilde{t} = 8$, using the same parameters as in a)-b). Value and errors correspond to the mean and standard deviation over 50 repetitions.
}\label{fig2}
\end{figure*}

\begin{figure*}[t]
\centering
\includegraphics[]{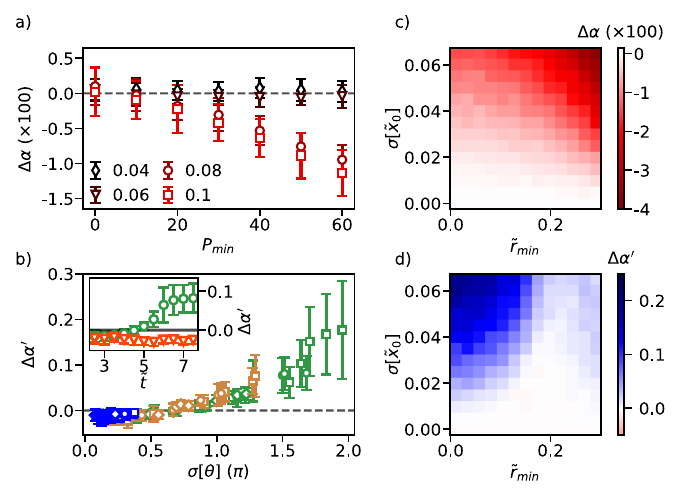}
\caption{a) $\Delta \alpha$ as a function of the
minimum annihilation probability $P_{min}$ considered in the fit at $\tilde{t} = 8$, using the same parameters as in Fig.~\ref{fig2}. Different colors and symbols represent different values of $\alpha$ (see legend). 
b) $\Delta \alpha '$ as a function of the standard deviation of the angular position for the same set of points in Fig.~\ref{fig2}b. 
The inset shows the comparison of the obtained $\Delta \alpha '$ for $\alpha ' = 0.2$, $\alpha = 0.08$ by estimating the average value of the angular position with the circular mean or directly as an average of the evolution of Eq.~\eqref{eq:theta}, shown as green circles and orange triangles, respectively.
c)-d) $\Delta \alpha$ and $\Delta \alpha '$ as functions of the initial standard deviation $\sigma[\tilde{x}_0] = \sigma[\tilde{y}_0]$ and of the annihilation threshold for the radial position $\tilde{r}_{min}$, for $\alpha = 0.08$ and $\alpha ' = 0.2$ at $\tilde{t} = 8$. 
}\label{fig3}
\end{figure*}

\section{Methods}

The dynamics of vortices in superfluids can be described using the dissipative point vortex model (DPVM). In this model, the effect of mutual friction on the vortex dynamics is described by the two mutual friction coefficients $\alpha$ and $\alpha '$ \cite{schwarz1985three,schwarz1988three,kopnin2002vortex,Sonin_magnusForce}. 
In the case of two-dimensional vortex dynamics, the velocity of a quantized vortex, $\vec{v}_L$, with circulation 
$ \vec{\kappa} = \sigma h / m \hat{z}$ (where $m$ is the mass of the superfluid constituent, and $\sigma=\pm1$ is the vortex circulation direction), moving in the $\hat{x}-\hat{y}$ plane is given by:
\begin{equation}
\label{VortexVelocity}
\vec{v}_L = \vec{v}_s + \alpha'(\vec{v}_n  - \vec{v}_s) + \alpha \sigma\hat z \times (\vec{v}_n  - \vec{v}_s).
\end{equation} 
Here, $\vec{v}_n$ and $\vec{v}_s$ represent the velocities of the normal and superfluid components in the two-fluid model, respectively.

In our recent study \cite{grani2025mutual}, we investigated the dynamics of vortices in a strongly interacting Fermi superfluid in a minimal configuration consisting of a single vortex–antivortex pair confined in a disk-shaped geometry, homogeneous in the $\hat{x}-\hat{y}$ plane. Due to the limited extent along the $\hat{z}$ direction, the vortex dynamics is treated as effectively two-dimensional.
By pinning the antivortex ($\sigma = -1$) at the center of the disk, we observe the motion of the mobile vortex, initially placed at position $\vec{r}(t=0) = (x_0, y_0) = (r_0,\theta_0)$.
The free vortex (with $\sigma = 1$) moves under the velocity field produced by the central one. To account for boundary effects, we include the additional contribution of the image antivortex located at $\vec{r}_{im} = \frac{R^2}{|r|^2} \vec{r}$ for the off-center vortex \cite{Imaginary_vortex_Disk}. Here, $\vec{r}$ is the position of the off-center vortex with respect to the center of the disk and $R$ is the radius of the disk.
As a result, the vortex at position $\vec{r}$ is subject to the superfluid velocity field given by: 
\begin{align}
 \vec{v}_s(\tilde{r}) &= - \frac{\kappa}{2\pi R}\frac{1}{\tilde{r}}\left( \frac{1-2\tilde{r}^2}{1-\tilde{r}^2} \right) \hat{\theta},
\label{superfluidvel}
\end{align}
with $\tilde{r} = \frac{r}{R}$.
Assuming the normal fluid is at rest in the laboratory frame ($\vec{v}_n = 0$), Eq.~\eqref{VortexVelocity} can be solved analytically for this velocity field. The resulting trajectory of the vortex with initial position $(\tilde{r}_0,\theta_0)$, when $\alpha \neq 0$, is:
\begin{align}
 \tilde{r}(\tilde{t}) &= \sqrt{\frac{1}{2}-\frac{1}{2}W\left(\left[1-2\tilde{r}_0^2\right]e^{1-2\tilde{r}_0^2+8\alpha\tilde{t}}\right)}, \label{eq:r}\\
 \theta(\tilde{t}) &= \theta_0 + \frac{1-\alpha'}{\alpha } \log \left(\frac{\tilde{r}(\tilde{t})}{\tilde{r}_0}\right), \label{eq:theta}
\end{align}
with dimensionless time $\tilde{t} = t/t_0$ where $t_0 = R^2/\kappa$ is the natural time scale of the system and
$W(x)$ denotes the Lambert W-Function. 

To evaluate the effect of the reproducibility of the initial conditions, we replicate the experimental procedure by studying the time evolution of vortex coordinates under DPVM, introducing fluctuations in the initial positions that are consistent with our experimental precision.
We note that our approach, tracking the deterministic dynamics of vortices starting from noisy initial conditions, is fundamentally different from that adopted in Refs. \cite{simjanovski2024shear,neely2024melting,reeves2022turbulent,mehdi2023mutual}, where stochastic noise is instead introduced during the time evolution of the vortex trajectories.
The results are then analyzed with the same procedure applied to the experimental data to extract the values of the mutual friction coefficients. Specifically, for each evolution time, we compute the average and standard deviation of the mean of the radial and azimuthal coordinates from 30 independent initial conditions normally distributed with standard deviations $\sigma [\tilde{x}_0] = \sigma[\tilde{y}_0]$ around the mean value ($\tilde{x}_0$,$\tilde{y}_0$) = ($\tilde{r}_0$,$\theta_0$). As for the experimental case, the angular position of each point is obtained from $\theta = \arctan(\tilde{y}/\tilde{x})$, with $\tilde{y} = y/R$ and $\tilde{x}= x/R$ being the normalized positions. The corresponding mean and standard deviation are calculated using the circular mean and circular standard deviation, respectively, as implemented in the SciPy library \cite{virtanen2020scipy,mardia2009directional}. The obtained values are then unwrapped, i.e., corrected for $2 \pi$ jumps in order to translate the values from the interval (-$\pi$, $\pi$] to ($-\infty$,$\pi$] and to ensure continuity of $\theta$. The mutual friction coefficients are obtained by fitting according to Eqs.~\eqref{eq:r}-\eqref{eq:theta} the resulting average positions at equally spaced time intervals, separated by $\Delta\tilde{t}$. This simulated experiment is repeated 50 times to obtain average values and standard deviations for the estimated mutual friction coefficients.

Fig.~\ref{fig1}a shows the time evolution of vortex positions obtained from the DPVM. As in the experimental case \cite{grani2025mutual}, small fluctuations in the initial positions lead to significantly different dynamical evolutions.  
As shown in Fig.~\ref{fig1}b, the measurement of the angular position (blue points) is confined to the interval ($-\pi$, $\pi$], which can differ by a multiple of $2\pi$ from the true value (yellow circles) obtained by Eq.~\eqref{eq:theta}. The use of the circular mean and the consequent unwrapping of the data overcome this issue (see red squares).
Considering Eq.~\eqref{VortexVelocity}, both the radial and angular velocity of the vortex are proportional to $v_s$.
Due to the nonlinearity of the superfluid velocity as a function of $\tilde{r}$, as given by Eq.~\eqref{superfluidvel}, the distributions of the vortex radial and azimuthal positions are expected to evolve asymmetrically, unbalancing in the direction of points with increased superfluid velocity respect to the central value (Fig.~\ref{fig1}c-d). 
As a result, the estimated average values of the radial and azimuthal positions tend to shift relative to the evolution of the initial distribution’s central value (indicated by the vertical lines in Fig.~\ref{fig1}c-d), leading to an overestimation of $\alpha$ and an underestimation of $\alpha'$ in the fitting procedure.
Notably, this nonlinear evolution is particularly evident in the distribution of azimuthal positions over time (see Fig.~\ref{fig1}d).
Moreover, this distribution exhibits significant broadening and, at long times, the angular positions can eventually differ by more than $2\pi$. 
However, points with angular positions differing by $2\pi$ 
at a given time are indistinguishable. 
As a consequence, over long times, the values obtained from the circular mean may represent an overestimation of the true mean value and lead to 
an overestimation of the value of $\alpha'$.
Finally, annihilation events, which in the DPVM time evolution correspond to a vanishing value of the intervortex distance, can bias the estimation of both the radial and angular positions, since these are more likely to occur for vortices with smaller initial radii \cite{kwon2021sound}. This effect leads to an overestimation of the radial position and an underestimation of the angular displacement, resulting in a reduced value of the observed $\alpha$ and an increased value of $\alpha'$.
To mitigate the impact of annihilation events, we restrict the analysis to time intervals during which the annihilation probability remains below the threshold value $P_{min}$.
Unless specified, we set $P_{min} = 30\%$ to remain consistent with the treatment of the experimental dataset.
In real systems, the minimum distance between two vortices before annihilation is characterized by a finite value $r_{\text{min}}$. This value is proportional to the size of the vortices $\xi_v$, on the order of the healing length \cite{ruostekoski2005engineering,jones1982motions} in bosonic systems, and to the inverse of the Fermi wave number $k_F^{-1}$ \cite{sensarma2006vortices,bulgac2003vortex,chien2006ground,simonucci2013temperature} in fermionic superfluids.
Moreover, depinning events of the central vortex may disrupt the vortex configuration at even larger radial distances.
To account for these effects in our analysis, we introduce a threshold value, $\tilde{r}_{min}$, for the normalized distance $\tilde{r}$, below which the configuration is considered annihilated.

\section{Results}\label{sec2}

Given that the effects are expected to compound over time, we study the temporal evolution of the deduced values of $\alpha$ and $\alpha '$.
Fig.~\ref{fig2} shows the results for $\tilde{r}_0=0.4$,
$\sigma [\tilde{x}_0] = \sigma[\tilde{y}_0] = 0.03 $, $\Delta \tilde{t} = 0.25$ and $\tilde{r}_{min} = 0$. 
As shown in Fig.~\ref{fig2}a, for long enough times, the value of $\alpha$ is underestimated.
We attribute this discrepancy to annihilation events, and, for the chosen $P_{min}$, the deviation from the real value is comparable to the error bars.
For higher values of $\alpha$, this effect manifests at shorter times, due to the faster radial decrease in this regime.
Instead, for $\alpha '$ (Fig.~\ref{fig2}b),
we observe that at short times its value is slightly underestimated, while at longer times it is significantly overestimated. 
We attribute the first effect to the nonlinear evolution of the angular distribution, and the second to the indistinguishability of points separated by more than $2\pi$ in the distribution.
The value of $\alpha'$ does not affect the deviation of $\alpha$, $\Delta \alpha$ (Fig.~\ref{fig2}c), as the latter depends solely on the radial time evolution.
On the contrary, as shown in Fig.~\ref{fig2}d, the deviation $\Delta \alpha '$ is less pronounced for higher values of $\alpha '$, as the reduced vortex velocity leads to a slower broadening of the distribution.

\begin{figure*}[th]
\centering
\includegraphics[]{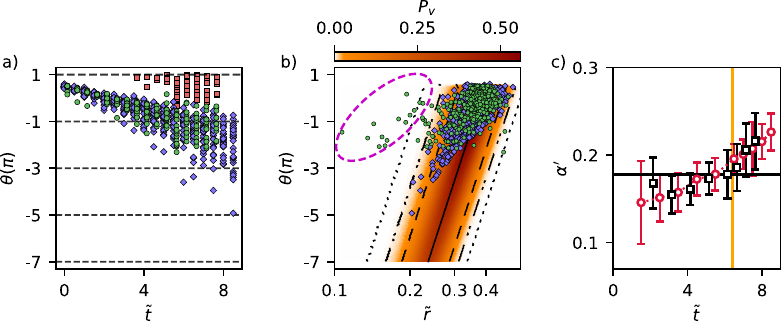}
\caption{a) Distribution of the angular position as a function of time. b) Relation between the observed angular and radial positions. 
Green points represent the experimentally observed distribution transposed in an interval of $\pm \pi$ around the mean value, with red squares being the observed experimental data, when they differ from the green ones, obtained as $\arctan(y/x)$.
Violet diamonds are the values obtained from DPVM (Eq.~\eqref{eq:theta}) with similar conditions. The orange colormap is a normalized density map obtained from $10^5$ DPVM time evolutions. Continuous, dashed, dot-dashed and dotted lines represent the value of the mean, $2\sigma$, $3\sigma$, $4\sigma$ of the radial distribution at a given value of $\theta$, respectively, with $\sigma$ being the standard deviation. The magenta dashed ellipse marks points out of the distribution for $\gtrsim 4\sigma$.
c) Measured $\alpha'$ as a function of time in the experimental case (black squares) without considering any annihilation threshold, and for the DPVM time evolution (red circles). Vertical orange line sets the maximum time considered in the experiment for the fitting, taking into account the annihilation threshold. The horizontal black line is the resulting fitted value and the value used in the DPVM time evolution analysis.}\label{fig4}
\end{figure*}

Fig.~\ref{fig3}a shows the deviation in $\alpha$ as a function of the minimal annihilation probability considered in the fit $P_{min}$. By increasing this value the deviation becomes more important for high values of $\alpha$, as expected from our interpretation of the origin of this effect in terms of annihilation events.
The interpretation of the observed increase in $\alpha '$ is highlighted in Fig~\ref{fig3}b, where the deviation of $\alpha '$ is displayed for the points presented in Fig.~\ref{fig2}b, as a function of the standard deviation of the angular position calculated directly from the time evolution of Eq.~\eqref{eq:theta}.
The observed trend suggests that the growth of $\Delta \alpha'$ is closely related to the spreading of the angular distribution.
Moreover, we observe that the overestimation at long times vanishes when the mean value of the azimuthal position is directly computed from the angular evolution as obtained from Eq.~\eqref{eq:theta}, i.e., when angles differing by more than $2\pi$ are distinguishable, consistently  with our interpretation (see inset).

Fig.~\ref{fig3}c-d show the effect of the value of the initial fluctuations $\sigma[\tilde{x}_0] = \sigma[\tilde{y}_0]$ and the minimum annihilation distance $\tilde{r}_{min}$ on $\Delta \alpha$ and $\Delta \alpha '$ for specific values of $\alpha$ and $\alpha'$. Deviations are more pronounced for larger initial position fluctuations. Furthermore, the deviation in $\alpha'$ decreases as $r_{min}$ increases. This behavior is attributed to a reduction in the maximum fitting time, as annihilation events are more likely, reducing the maximum spread reached in the angular position. In contrast, increasing $r_{min}$ leads to a larger $\Delta \alpha$, as the probability of having a vortex dipole decreases more rapidly, thereby having a stronger impact on the fitting.

For most of our experimental realizations, reported in Ref.~\cite{grani2025mutual}, the deviations described above are not detected in the time evolution of the mutual friction coefficients. Compatibly, DPVM time evolution for the experimental parameters predicts deviations smaller than the measured error bars and the experimental noise.
Fig.~\ref{fig4} illustrates a specific case in which the value of $\alpha'$ appears to be overestimated at long times, due to a particularly noisy initial condition. 
To investigate this effect, we compare the experimentally observed angular position distribution, within an interval of $\pm \pi$ around the estimated average value at a given time, with the corresponding distribution obtained from DPVM simulations under similar conditions (Fig.\ref{fig4}a).
At long times, the latter spans a range exceeding $2\pi$, while the experimental data begin to deviate from the DPVM distribution, suggesting that some angular positions may be inaccurately estimated.
Eq.~\eqref{eq:theta} provides a relation between $\theta(t)$ and $\tilde{r}(t)$. As a result, badly-estimated angular values manifest as points that deviate from the expected DPVM distribution (green points in the top left region in Fig.~\ref{fig4}b, enclosed by the magenta dotted ellipse), representing a way to distinguish unreliable experimental points.
As time proceeds, we observe an increasing trend in the experimentally extracted value of $\alpha'$ (Fig.~\ref{fig4}c), when the fit is performed without applying any annihilation threshold. This trend is reproduced by the DPVM simulation results. 
However, by imposing the fitting threshold for annihilation of $P_{\text{min}} = 30\%$, the fitting interval is truncated at an earlier time (indicated by the orange vertical line), thereby preventing the overestimation of $\alpha'$ in this experimental run.

\section{Conclusion}\label{sec13}

We have investigated how fluctuations in the initial vortex position affect the estimation of mutual friction coefficients, demonstrating that the measured values can be systematically biased. 
Our analysis shows that such biases can be identified through the time evolution of the measured coefficients, and we have discussed possible physical origins for these effects. Moreover, we provide an experimental evidence of a possible overestimation of $\alpha '$. The potential deviations in the mutual friction values reported in \cite{grani2025mutual} were considered during our data analysis by applying the methodology presented in this work to the specific experimental parameters. For the typical timescale of our experimental data, this revealed that any systematic biases introduced by vortex position fluctuations are smaller than the experimental error bars. 
Given that our experimental configuration is broadly applicable to the study of mutual friction across different superfluid regimes, our findings offer important insights for improving the accuracy of such measurements. These effects are expected to be especially relevant in experiments involving BCS superfluids, where the mutual friction parameter $\alpha$ is expected to be larger \cite{kwon2021sound} than in our observations \cite{grani2025mutual}.
The ability to accurately track vortex motion could be used as a powerful tool to investigate vortex mass in BCS superfluids \cite{richaud2024dynamical,levrouw2025vortex, caldara2024suppression} and binary Bose-Einstein condensates \cite{bellettini2023relative,caldara2023massive}, as well as to reveal disordered phase structures in spin-imbalanced BCS systems \cite{magierski2024quantum} and to estimate the superfluid fraction \cite{richaud2025probing}. The type of analysis presented in this work can be extended to these scenarios to evaluate the impact on measurable quantities, or to determine the precision required in the initial conditions to observe these effects in experimental vortex trajectories affected by noise.

\section*{Acknowledgements}
We thank the Quantum Gases group at LENS for fruitful discussions.
G.R. and G.D.P. acknowledge financial support from the PNRR MUR project PE0000023-NQSTI. G.R. acknowledges funding from the Italian Ministry of University and Research under the PRIN2017 project CEnTraL and the Project CNR-FOE-LENS-2024. The authors acknowledge support from the European Union - NextGenerationEU for the “Integrated Infrastructure initiative in Photonics and Quantum Sciences" - I-PHOQS [IR0000016, ID D2B8D520, CUP B53C22001750006]. The authors acknowledge funding from INFN through the RELAQS project. This publication has received funding under the Horizon Europe program HORIZON-CL4-2022-QUANTUM-02-SGA via project 101113690 (PASQuanS2.1) and by the European Community's Horizon 2020 research and innovation program under grant agreement n° 871124.

\section*{Competing interests}
The authors declare no competing interests.

\section*{Data availability}
The data that support the figures and the results within this paper are available from the corresponding author upon reasonable request.

\section*{Author contribution}
N.G. and G.R. conceived the study.
N.G. carry out and analyzed the DPVM time evolutions.  
All authors contributed to the interpretation of the results and to the writing of the manuscript.



\end{document}